\documentclass[aps,english,noeprint,longbibliography, twocolumn, 10pt, superscriptaddress, floatfix, prb]{revtex4-2}

\pdfoutput=1
\usepackage[utf8]{inputenc}
\usepackage[T1]{fontenc}
\usepackage{verbatim}
\usepackage[version=3]{mhchem}
\usepackage{natbib}
\usepackage{mathtools}
\usepackage{amsmath}
\usepackage{amssymb}
\usepackage{graphicx}
\usepackage{wasysym}
\usepackage{layouts}
\usepackage{siunitx}
\usepackage{longtable}
\usepackage{booktabs}
\usepackage{bm}
\usepackage{xcolor}
\usepackage[colorlinks, citecolor={blue!50!black}, urlcolor={blue!50!black}, linkcolor={red!50!black}]{hyperref}
\usepackage{bookmark}
\usepackage{tabularx}
\usepackage{microtype}
\usepackage{times}
\usepackage{babel}
\hypersetup{pdfauthor={Joseph A. M. Paddison, Matthew J.
Cliffe},pdftitle={Discovering classical spin liquids by topological search of high symmetry nets
}}

\DeclarePairedDelimiter\abs{\lvert}{\rvert}
\DeclarePairedDelimiter\norm{\lVert}{\rVert}

\makeatletter
\let\oldabs\abs
\def\abs{\@ifstar{\oldabs}{\oldabs*}}
\let\oldnorm\norm
\def\norm{\@ifstar{\oldnorm}{\oldnorm*}}
\makeatother

\newcolumntype{L}[1]{>{\raggedright\arraybackslash}p{#1}}
\newcolumntype{C}[1]{>{\centering\arraybackslash}p{#1}}
\newcolumntype{R}[1]{>{\raggedleft\arraybackslash}p{#1}}
\renewcommand{\citep}{\cite}

\begin{document}

\title{Discovering classical spin liquids by topological search of high symmetry nets}

\author{Joseph A. M. Paddison}
\email[Electronic address: ]{paddisonja@ornl.gov}
\affiliation{Neutron Scattering Division, Oak Ridge
National Laboratory, Oak Ridge, Tennessee 37831, USA}
\author{Matthew J. Cliffe}
\email[Electronic address: ]{matthew.cliffe@nottingham.ac.uk}
\affiliation{School of Chemistry, University Park, Nottingham, NG7 2RD,
United Kingdom}

\date{\today}

\begin{abstract}
  \begin{center}
    \textbf{Abstract}
  \end{center}
Spin liquids are a paradigmatic example of a non-trivial state of
matter, and the search for new spin liquids is a key direction in physics, chemistry, and materials science. Geometrical frustration---where the
geometry of the net that the spins occupy precludes the generation of
a simple ordered state---is a particularly
fruitful way to generate these intrinsically disordered states. A
particular focus has been on a handful of high symmetry nets. There are, however,
many three-dimensional nets, each of which has the potential to form
unique states. In this paper, we investigate
the high symmetry nets---those which are both vertex- and
edge-transitive---for the simplest possible interaction sets:
nearest-neighbor couplings of antiferromagnetic Heisenberg and Ising spins.
While the well-known \textbf{crs} (pyrochlore) net is the only nearest-neighbor Heisenberg
antiferromagnet which does not order, we identify two new frustrated nets (\textbf{lcx} and \textbf{thp})
that possess finite temperature Heisenberg spin-liquid states with strongly
suppressed magnetic ordering and non-collinear ground states. With Ising spins, we identify three new classical
spin liquids that do not order down to \(T/J = 0.01\). We highlight
materials that contain these high symmetry nets, and which could, if substituted
with appropriate magnetic ions, potentially host these unusual states. Our
systematic survey will guide searches for novel magnetic phases.
\end{abstract}

\flushbottom
\maketitle

\hypertarget{introduction}{%
\section{Introduction}\label{introduction}}

Frustrated magnets, where the combination of geometry and interactions
suppress the emergence of long-range ordered states, are rich in
unconventional magnetic states: from complex spin
textures\citep{henleyCoulombPhaseFrustrated2010,paddisonCubicDoublePerovskites2024}
to spin
liquids.\citep{ramirezStronglyGeometricallyFrustrated1994, balentsSpinLiquidsFrustrated2010, diepFrustratedSpinSystems2013, lacroixIntroductionFrustratedMagnetism2011}
However, the existence of frustration implies a (near-)degeneracy of
states, and hence that the nets on which the spins are arranged must be
highly symmetric. Most research on frustrated magnets has focused on a
handful of the most symmetric nets: in two dimensions, the kagome
(\textbf{kgm})\citep{kanoAntiferromagnetismKagomeIsing1953, reimersOrderDisorderClassical1993}
and triangular
(\textbf{hxl})\citep{Wannier1950, kawamuraPhaseTransitionTwoDimensional1984}
nets, and in three, the pyrochlore (\textbf{crs})
net,\citep{moessnerLowtemperaturePropertiesClassical1998, moessnerPropertiesClassicalSpin1998}
where the degeneracy of states and hence the frustration is greatest.

Although we now understand much about the behavior of spins arranged on
these high symmetry nets, many other nets are less well understood, particularly in three dimensions. A diversity of topologies has been created in metal-organic framework (MOF)
crystals and enabled a range of functional properties, often using the pore network-topology to achieve chemical
separations\citep{Yaghi2003a, bhattTopologyMeetsReticular2020}, but
increasingly also for realising novel magnetic nets.
\citep{bulledGeometricFrustrationTrillium2022} These results highlight that the
synthetic chemistry need not be a barrier to moving beyond
traditional nets. Determining whether a given net could host
frustration-derived unconventional magnetic states is a key step towards the discovery of such states in real materials, whether through targeted
`reticular' design of new magnets or through computational search of
known materials.\citep{meschkeSearchStructuralFeaturization2021}

\begin{figure}[b]
\centering
\includegraphics{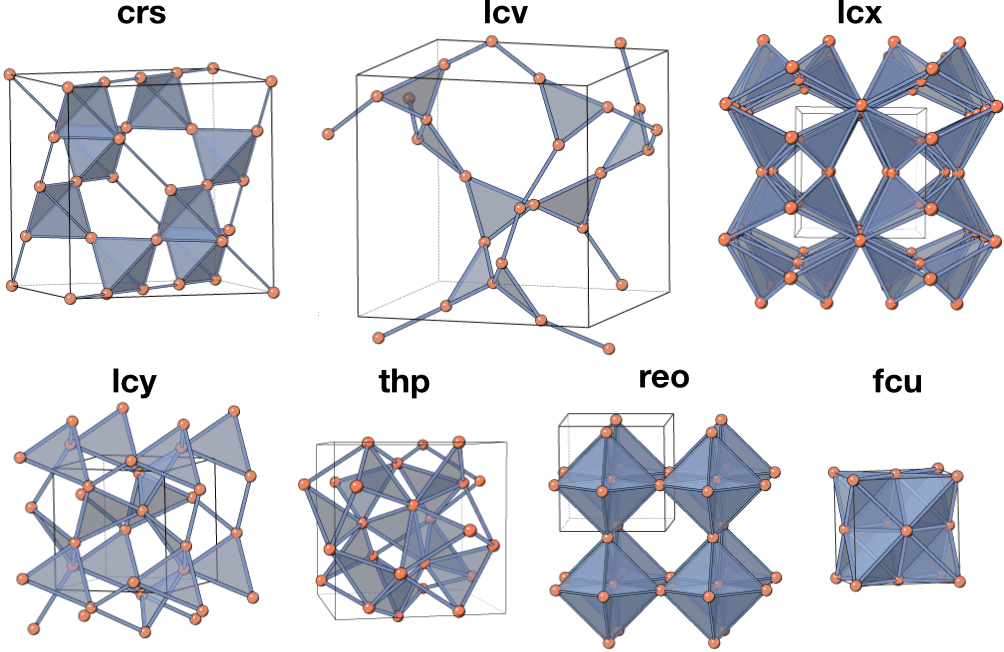}
\caption{Vertex- and edge-transitive frustrated nets in their high symmetry embeddings. For each net, the edges are the shortest vertex-vertex distance,
 except for \textbf{lcx} for which the edges are the next-shortest vertex-vertex distance.}
\label{fig:nets}
\end{figure}

The most important class of nets comprises those with only one symmetry
independent vertex (vertex-transitive) and only one symmetry independent
nearest-neighbor interaction (edge-transitive). Examples of these nets are shown in Fig.~\ref{fig:nets} and listed in Table~\ref{tbl:frust_net}. 
These are the only nets in
which a frustrated state can be achieved without fine-tuning, as all
other nets will have at least two symmetry-inequivalent
nearest-neighbor interactions. Each vertex and edge-transitive net can
generate an infinite number of other vertex and edge-transitive nets, by
connecting each vertex to its \(n\)th-neighbor rather than the nearest neighbor (\(n=1\));
however, these higher-order nets are rarely found in
real materials, as their highest symmetry realisations typically require
intersecting edges, very high degrees of interpenetration, very long
ligands, or very high coordination numbers. 

\footnotetext[1]{To the best of our knowledge there is no formal proof that these 20 nets represent the only possible vertex- and edge-transitive 3D nets with nearest-neighbor edges, but  extensive searches have found no others.\citep{delgado-friedrichsThreeperiodicTilingsNets2007,blatovTopologicalRelationsThreedimensional2007}} 

In this paper, we investigate the behavior of classical antiferromagnets of the 21 `simplest' vertex- and edge-transitive 3D nets. These nets have many different names {[}Table \ref{tbl:frust_net}{]},
and so to identify each net unambiguously we use the Reticular Chemistry
Structural Resource (RCSR) nomenclature, where each net is assigned a
three letter code written in bold.\citep{OKeeffe2008} 
We restrict our study to the vertex- and edge-transitive nets generated by connecting the vertices
nearest in space to each other in their highest symmetry
embedding (20 nets). We also consider the \textbf{lcx} net in which next-nearest
neighbor vertices are connected, because the nearest neighbor net is only 1
connected, \(r_1/r_2 = 0.82\), and it has been physically
realized. Further detail on these nets is available in the supporting information.\citep{delgado-friedrichsThreeperiodicTilingsNets2007, blatovTopologicalRelationsThreedimensional2007} 
Further, we focus on the simplest interactions, with a single antiferromagnetic interaction $J$ that couples nearest-neighbor spins only. We consider Heisenberg
\(E = J\sum_{\langle{i,j}\rangle} \mathbf{S}_{i}\cdot\mathbf{S}_{j}\) and Ising \(E = J \sum_{\langle{i,j}\rangle} S_i S_j\) models,
where $\mathbf{S}_{i}$ is a classical vector, \(S_i = \pm 1\), and the summation is over all nearest-neighbor pairs (next-nearest neighbor pairs for \textbf{lcx}). While more complex interactions
(\emph{e.g.}, bond-directional exchange\citep{Kitaev2006}) or multiple exchange
interactions \(J_n\)\citep{bergmanOrderbydisorderSpiralSpinliquid2007} can also yield interesting physics, in many cases
the Heisenberg interaction remains the largest and physically most relevant term.

We employ Monte Carlo simulation on large supercells to provide an overview of the magnetic properties for these high-symmetry nets, and
to calculate quantities that can be compared with experimental data for magnetic framework materials.
We provide several measures of the degree of frustration for each of these nets, including the `frustration parameter' accessible to bulk magnetic susceptibility measurements, and the magnetic residual entropy accessible to heat-capacity measurements. We calculate the magnetic ground states, which can be measured experimentally using neutron diffraction below the magnetic ordering temperature $T_\mathrm{N}$, and the spin-spin correlation functions in the spin-liquid (correlated paramagnetic) regime, which are accessible to neutron scattering measurements above $T_\mathrm{N}$. 
Our results reproduce established results for well-known nets such as \textbf{crs} (pyrochlore). Importantly, however, our survey also identifies two nets that have not previously been investigated, \textbf{lcx} and \textbf{thp}, which have significant frustration as Heisenberg antiferromagnets and adopt non-collinear magnetic ground states. Moreover, we find that five nets composed of corner-sharing triangles show no order down to \(T/J = 0.01\) when populated with Ising spins. These results will facilitate topology-guided synthesis and experimental identification of new magnetic materials that realize novel magnetic topologies.

\begin{table}

  \caption{Common names of the high symmetry topologies, coordination number $z$, and examples of
compounds with these topologies.
\label{tbl:frust_net}}
  
  \begin{tabular}{p{0.06\linewidth} p{0.05\linewidth}  >{\raggedright}p{0.28\linewidth}  p{0.38\linewidth}  p{0.13\linewidth} }
%  \topple
 & \emph{z}   &  Synonyms & Compounds & Space Group \\
  \midrule
  \textbf{crs} & 6 & pyrochlore, spinel B-site, cubic Laves &
Ho\(\subset\)\(\ce{Ho2Ti2O7}\),\citep{bramwellSpinIceState2001}
Cr\(\subset\)\(\ce{MgCr2O4}\)\citep{baiMagneticExcitationsClassical2019}
& \(F d \overline{3} m\) \\
\textbf{fcu} & 12 & face-centered cubic, rocksalt, double perovskite &
Mn\(\subset\)MnO,\citep{shullDetectionAntiferromagnetismNeutron1949}
Ir\(\subset\)\(\ce{K2IrCl6}\)\citep{khanCubicSymmetryMagnetic2019} &
\(F m \overline{3} m\) \\
\textbf{lcv} & 4 & hyperkagome, (half-)garnet &
Gd\(\subset\)\(\ce{Gd3Ga5O12}\),\citep{Paddison2008}
Ir\(\subset\)\(\ce{Na4Ir3O8}\)\citep{okamotoSpinLiquidStateHyperkagome2007}
& \(I4_132\) \\
\textbf{lcx} & 8 & next-nearest neighbor \(A15\) phase &
Cr\(\subset\)\ce{Cr3Si},
\citep{blaugherAtomicOrderingSuperconductivity1969}
Pt\(\subset\)\ce{NaPt3O4},\citep{waserStructureNaxPt3O41952}
Eu2\(\subset\)\ce{Eu8Ga18Ge30}\citep{salesStructuralMagneticThermal2001}
& \(Pm\overline{3}n\) \\
\textbf{lcy} & 6 & trillium
&
Mn\(\subset\)\(\ce{NaMn(HCO2)3}\),\citep{bulledGeometricFrustrationTrillium2022}
Mn\(\subset\)\(\ce{MnSi}\)\citep{pfleidererPartialOrderNonFermiliquid2004}
& \(P4_132\) \\
\textbf{reo} & 8 & octahedral, octochlore & Mn\(\subset\)\(\ce{Mn3ZnN}\)
\citep{fruchartMagneticStudiesMetallic1978} & \(Pm\overline{3}m\) \\
\textbf{thp} & 8 & thorium phosphide & U\(\subset\)\(\ce{U3X4}\), X=P,As,Sb,Bi\citep{wisniewskiSpinOrbitalMoments1999} &
\(I\overline{4}3d\) \\
  \bottomrule
  \end{tabular}
\end{table}

\hypertarget{results}{%
\section{Results}\label{results}}

To investigate the ground states of these models, we carried out Metropolis Monte
Carlo simulations on $10\times10\times10$ supercells (except where noted) with periodic boundary conditions, which were gradually cooled from a high temperature ($T/J=5$ to $100$). A Monte Carlo `move' involves proposing a single spin flip for Ising simulations, or an over-relaxation update followed by a random spin rotation for Heisenberg spin systems. At each temperature, the number of proposed moves $n_\mathrm{c}$ needed to decorrelate the supercell was estimated; simulations were typically run for $10n_\mathrm{c}$ for equilibration followed by $100$ to $4000n_\mathrm{c}$ for measurement. It was usually possible to decorrelate the supercells and maintain ergodicity, with the exception of the Ising \textbf{fcu} net, which rapidly freezes below $T_\mathrm{N}$.

\hypertarget{heisenberg}{%
\subsection{Heisenberg}\label{heisenberg}}

\begin{figure}
  \centering
  \includegraphics[scale=0.8]{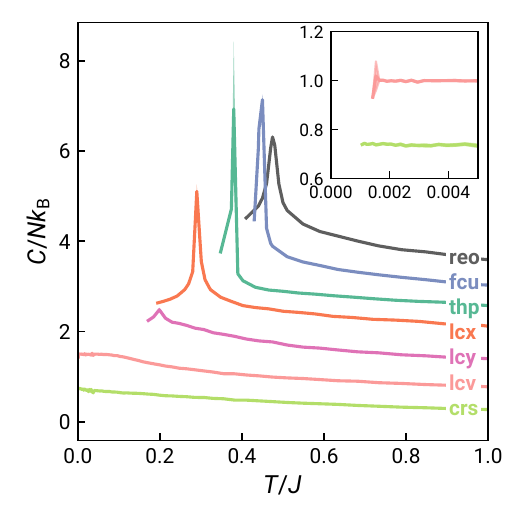}
  \caption{Heat capacity for the seven frustrated nets with nearest neighbor Heisenberg interactions derived from Monte Carlo simulations of $10\times10\times10$ supercells ($6\times6\times6$ supercells for \textbf{crs} and \textbf{lcv}). Curves are offset vertically for clarity. The inset shows the low-temperature heat capacity for \textbf{crs} and \textbf{lcv}, showing the low-temperature transition in \textbf{lcv} at $T/J\approx 0.002$.\cite{zhitomirskyOctupolarOrderingClassical2008} 
  }
  \label{fig:heisenberg_heat_cap}
\end{figure}

\begin{figure*}
\centering
\includegraphics{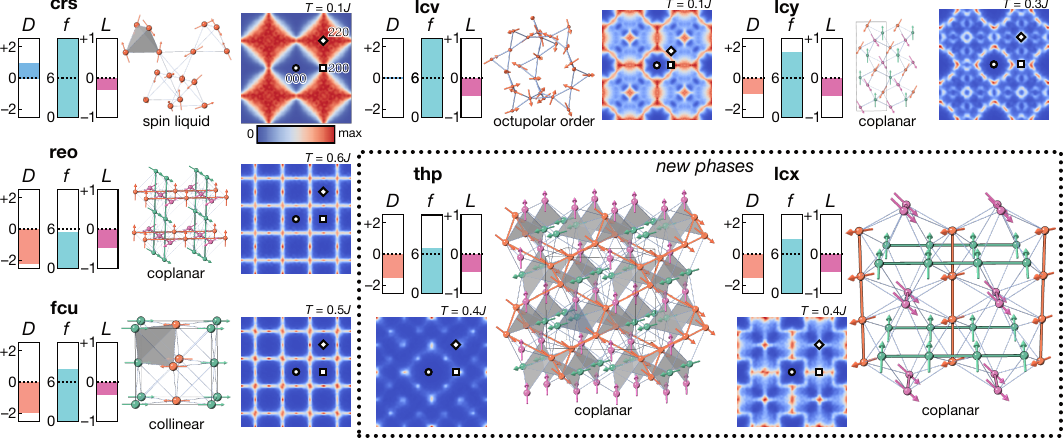}
\caption{Summary of properties of the frustrated nets with Heisenberg spins. For each net, we show (left) three measures of the degree of frustration, defined such that larger values indicate a higher degree of frustration: $D$ is the Maxwellian degrees of freedom defined in the text, $f = zJ/3T_\mathrm{N}$ is the frustration parameter, and $L=-E/E_\mathrm{B}$ is the Lacorre constraint function, where $E$ is the energy per spin at $T/J=0.1$. Structures shown are the ground states, with spin orientation indicated by arrows. The distinct substructures (for ordered phases) are indicated by different colors, and grey highlights show triangles or tetrahedra. Magnetic scattering $S(\mathbf{q})=\left|\sum_{i}\mathbf{S}_{i}\exp(\mathrm{i}\mathbf{q}\cdot\mathbf{r}_{i})\right|^{2}$ in the paramagnetic phase is shown in the $hk0$ plane, and symmetrized in $m\overline{3}m$ symmetry. Temperatures of diffuse scattering calculations are labelled in each panel ($T/J = 0.1$ for phases without antiferromagnetic ground states, or just above $T_\mathrm{N}$ for phases with antiferromagnetic ground states). The $(000)$ position is indicated by a circle, $(200)$ by a square, and $(220)$ by a diamond. A color bar indicating the intensity scale is shown below \textbf{crs}.}
\label{fig:heisenberg_summary}
\end{figure*}

\begin{table}

\caption{Ground states of the Heisenberg nearest-neighbor antiferromagnet on the high symmetry frustrated nets. The frustration parameter $f = zJ/3T_\mathrm{N}$, and the Lacorre constraint function \(L=-{E}/{E_\mathrm{B}}\), where $E$ is the energy per spin at $T/J=0.1$. \label{tbl:heisenberg}}

\begin{tabular}{llrrrll}
\toprule
  &  $T_\mathrm{N}/J$ & $f$ & $L$ & $D$ & Ground state & $\mathbf{k}$-vector \\
\midrule
\textbf{crs} & \textless0.001 & \textgreater2000 &  $-0.31$ & 1 & Spin
liquid \citep{moessnerLowtemperaturePropertiesClassical1998, moessnerPropertiesClassicalSpin1998}
& - \\
\textbf{fcu} & 0.45 & 8.9 & $-0.32$ & $-2$ & Collinear order
\citep{gvozdikovaMonteCarloStudy2005,oguchiSpinWaveTheory1985, schickGroundstateSelectionMagnon2022} &
(1,0,0)$^\dag$\\
\textbf{lcv} & 0.002 & 700 & $-0.45$ & 0 & Octupolar order
\citep{zhitomirskyOctupolarOrderingClassical2008} & - \\
\textbf{lcx} & 0.29 & 9.2 & $-0.48$ & $-1.5$ & 120\(^\circ\) order\(^\ast\) &
(0,0,0) \\
\textbf{lcy} & 0.20 & 10 & $-0.46$ & $-1$ & 120\(^\circ\)
order \citep{isakovFatePartialOrder2008, hopkinsonGeometricFrustrationInherent2006}
& (\(\frac{1}{3}\),0,0) \\
\textbf{reo} & 0.48 & 5.6 & $-0.47$ & $-2.25$ & 120\(^\circ\) order 
\citep{sklanNonplanarGroundStates2013} & (0,0,0) \\
\textbf{thp} & 0.38 & 7.0 & $-0.47$ & $-2$ & 120\(^\circ\) order\(^\ast\) &
(0,0,0) \\
\bottomrule
\end{tabular}

\(\ast\)
This work. $\dag$ $\mathbf{k}=(1,0,0)$ order forms at $T_\mathrm{N}$ as it is favored by thermal fluctuations, but the ground-state selection between $(1,0,0)$ and $(1,\frac{1}{2},0)$ order is subtle.\citep{schickGroundstateSelectionMagnon2022}
\end{table}

Of the 21 high symmetry nets, 14 are
bipartite (see Supplementary Table~1 and Supplementary Figure~1).\citep{delgadofriedrichsThreeperiodicNetsTilings2003} Since the bipartite nets order on cooling into nearest neighbor antiferromagnetic
states, we do not consider them further. Seven nets, which all have cubic symmetry, are non-bipartite:
\textbf{crs} (pyrochlore), \textbf{fcu} (face-centered cubic),
\textbf{lcv} (hyperkagome), \textbf{lcx} (lattice complex X),
\textbf{lcy} (trillium), \textbf{reo} (octochlore) and \textbf{thp}
(thorium phosphide). These nets all are frustrated and hence the
magnetic ordering temperature is significantly suppressed, as indicated by the frustration parameter $f = zJ/3T_\mathrm{N}$, where $z$ is number of nearest neighbors. Their ground state energies $E$ are also significantly higher than the summed absolute values of all pairwise energies $E_\mathrm{B}$, as quantified by the ratio \(L=-{E}/{E_\mathrm{B}}\), which is called the Lacorre constraint function and varies between $-1$ (no frustration) and $+1$ (maximal frustration).\citep{lacorreConstraintFunctionsAttempt1987}

Results of our Monte Carlo simulations are
summarized in Figure~\ref{fig:heisenberg_summary} and Table~\ref{tbl:heisenberg}. All seven nets
ordered above \(T/J = 0.1\) except \textbf{crs} (pyrochlore) and
\textbf{lcv} (hyperkagome), as previously reported [Fig.~\ref{fig:heisenberg_heat_cap}].\citep{moessnerLowtemperaturePropertiesClassical1998, moessnerPropertiesClassicalSpin1998, zhitomirskyOctupolarOrderingClassical2008}
We note that \textbf{lcv} also undergoes a transition at \(T/J\approx 0.002\) into a coplanar ground state with an octupolar order parameter [Fig.~\ref{fig:heisenberg_heat_cap} and Supplementary Figure~2]. \citep{hopkinsonClassicalAntiferromagnetHyperkagome2007,zhitomirskyOctupolarOrderingClassical2008} Magnetic correlation functions are shown in Supplementary Figures~2 and 3 to demonstrate ordering (or its absence) for each net.
Maxwellian counting arguments\citep{moessnerLowtemperaturePropertiesClassical1998,moessnerMagnetsStrongGeometric2001} can be used to predict the degrees of freedom per spin and hence the level of frustration: for a net comprising polyhedra each containing $q$ spins with $n$ components ($n=3$ for Heisenberg spins) connected to $b$ other polyhedra, assuming independence of constraints, the degree of freedom per spin $ D = q(n-1)/b-n$. The value of $D$ predicts that \textbf{crs} and \textbf{lcv} are the two least constrained, and hence most frustrated, nets {[}Fig. \ref{fig:heisenberg_summary}{]}. For the five nets that order above \(T/J = 0.1\), two have been previously investigated using Monte Carlo simulations and we reproduce these results, finding collinear order in
\textbf{fcu}\citep{gvozdikovaMonteCarloStudy2005,oguchiSpinWaveTheory1985, schickGroundstateSelectionMagnon2022}
and 120\(^\circ\) coplanar order in
\textbf{lcy}.\citep{isakovFatePartialOrder2008, hopkinsonGeometricFrustrationInherent2006} The \textbf{reo} net has a large degeneracy of 120$^{\circ}$ ordered states at $T=0$;\citep{sklanNonplanarGroundStates2013} however, our Monte Carlo simulations indicate that coplanar 120\(^\circ\) order is selected by thermal fluctuations (`order-by-disorder' mechanism\citep{villainOrderEffectDisorder1980}).
Two nets have not previously been investigated as frustrated magnets,
\textbf{lcx} and \textbf{thp}, and we discuss them further below.

The \textbf{thp} net consists of corner-sharing triangles, and each vertex is
connected to four different triangles, giving a coordination number of
eight. This topology differs from the well-known corner-sharing triangular nets \textbf{kgm} and \textbf{lcv},
where each vertex is connected to two triangles, and from \textbf{lcy}, where
each vertex connects to three triangles.
The smallest non-triangle cycle in a \textbf{thp} structure is four atoms.
Despite its complex appearance, \textbf{thp} is the underlying net in a
number of materials; \emph{e.g.}, it is the net formed by connecting Th
atoms in \ce{Th3P4}---hence the name---and the actinide (Ac) cations in
the broader actinide pnictides, Ac\(_3\)X\(_4\), X\,=\,P, As, Sb or Bi.
The most thoroughly characterized magnets with the \textbf{thp} net are
the uranium analogues \ce{U3X4}, X\,=\,P, As, Sb or
Bi.\citep{wisniewskiSpinOrbitalMoments1999} These are metallic
ferromagnets with strong local anisotropy, due to the presence of large crystal field effects and RKKY type interactions, which produce non-collinear magnetism in \ce{U3P4} and
\ce{U3As4}.\citep{burletNoncollinearMagneticStructure1981} These effects are likely to be present in other actinide magnets with the \textbf{thp} net. The
\textbf{thp} net is closely related to the edge-transitive, but not
vertex-transitive, \textbf{ctn} net, which consists of 3- and 4-
coordinate vertices, and is widely found in MOFs and
COFs,\citep{Zhang2013e, el-kaderiDesignedSynthesis3D2007} as it is the
net formed by considering only the 4-coordinate vertices. The \textbf{thp}
net therefore describes the Si net of the \textbf{ctn}-topology
mineral eulytite (\ce{Bi4(SiO4)3})
\citep{fischerComparisonNeutronDiffraction1982} as well as the metal
atom sites in the Zn-MOF MOAAF-1.\citep{manosNewZn2Metal2012}

We found that the ground state of the Heisenberg \textbf{thp}
antiferromagnet is a coplanar 120\(^{\circ}\) structure in which
the three spin orientations occupy three distinct substructures, each
substructure with the \textbf{dia} topology. This ordering breaks the
centre of symmetry, generating a chiral structure: the highest symmetry
spin structure it can adopt, with all spins lying perpendicular to a
\(\langle 111\rangle_\mathrm{cubic}\) direction, has \(R3\)
space group symmetry {[}Fig. \ref{fig:heisenberg_summary}{]}.

The \textbf{lcx} net, like the \textbf{thp}, is eight-coordinate and
assembled from corner-sharing triangles in which each vertex belongs to
four triangles. As previously discussed, unlike the other twenty nets
considered, the shortest distance between vertices does not form an edge of
the net. If only this shortest distance were to be used as an
edge, the primitive rod packing (in which rods are stacked in three
orthogonal directions) would be
obtained.\citep{okeeffeRodPackingsCrystal1977} There are a number of
materials which contain the \textbf{lcx} net, including Pt
atoms in \ce{Pt3O4},\citep{mullerFormationStabilityPlatinum1968} and
MOFs derived from the \textbf{pto} net (which is 3- and 4-coordinated), such as the \ce{Cu2(CO2)4} dimers in MOF-14.
\citep{chenInterwovenMetalOrganicFramework2001} We are unaware of any
reports of magnetic properties of a material where magnetic ions populate the \textbf{lcx} net, although the next-nearest
neighbor connections for Eu2 in ferromagnetic \ce{Eu8Ga16Ge30} would do so. 
\citep{salesStructuralMagneticThermal2001} Like the \textbf{thp} net,
the ground state of the Heisenberg \textbf{lcx} antiferromagnet is a
non-collinear 120\(^{\circ}\) structure, in which the three spin
orientations occupy three distinct substructures, in this case consisting
of the 1D orthogonal chains found in the nearest neighbor net. The
maximal symmetry of this ordered structure is \(R\overline{3}\), again
when the spins lie perpendicular to a
\(\langle 111\rangle_\mathrm{cubic}\) direction {[}Fig.
\ref{fig:heisenberg_summary}{]}.

For all these nets, we find that above \(T_\mathrm{N}\), there is
significant structured magnetic diffuse scattering, indicative of strong
spin-spin correlations in a cooperative paramagnet (classical spin liquid) regime {[}Fig. \ref{fig:heisenberg_summary}{]}.
Qualitatively, the degree of structure found in the diffuse scattering
is inversely related to the frustration: the more closely the diffuse
scattering resembles well-defined Bragg peaks, the less frustrated the
system. The greater degree of structure in the \textbf{lcx} and \textbf{thp}
diffuse scattering is consistent with the weaker frustration in these nets, compared with \textbf{crs} and \textbf{lcv}.

\begin{figure}
\centering
\includegraphics{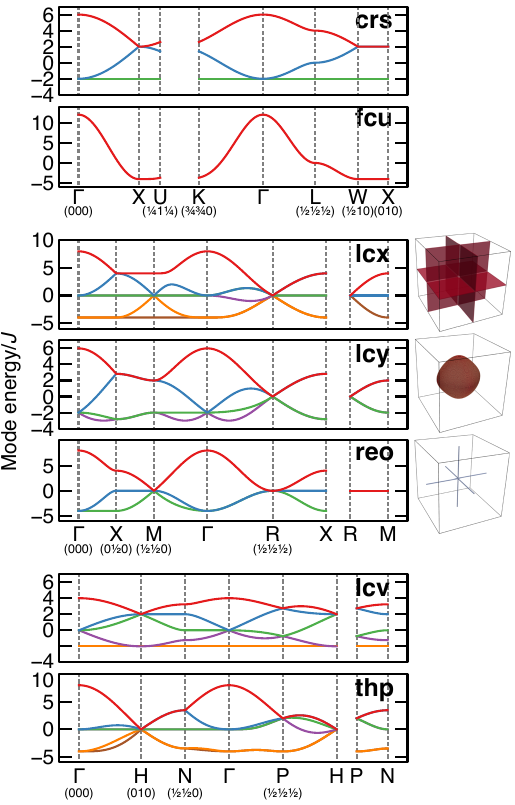}
\caption{Dispersion of eigenvalues of the interaction matrix $\mathsf{J}(\mathbf{q})$ for frustrated nets. For \textbf{lcx}, \textbf{lcy}, and \textbf{reo}, which possess primitive lattice-centering, the Brillouin zone is also shown (right) with the minimum-energy surface highlighted.}
\label{fig:k_paths}
\end{figure}

Further insight into the frustration of these nets is provided by the interaction matrix $\mathsf{J}(\mathbf{q})$.\citep{reimersMeanfieldApproachMagnetic1991,canalsMeanfieldStudyDisordered2000} For a net with $N$ atoms in its primitive unit cell, $\mathsf{J}(\mathbf{q})$ is an $N\times N$ matrix with elements $J_{ij}(\mathbf{q}) = \sum_{\mathbf{R}} J_{ij}(\mathbf{R})\exp(\mathrm{i}\mathbf{q}\cdot\mathbf{R})$, where $J_{ij}(\mathbf{R})\in \{0,J\}$ is the interaction between atom $i$ in a unit cell at the origin, and atom $j$ in the unit cell at lattice vector $\mathbf{R}$. In a mean-field approximation, the propagation vector of the ordered state is the wavevector (or set of wavevectors) at which the minimum eigenvalue of $\mathsf{J}(\mathbf{q})$ is located.\citep{reimersMeanfieldApproachMagnetic1991} 
For a net without frustration, the propagation vectors form a small set of symmetry-related points. By contrast, in frustrated systems, there is a degeneracy of propagation vectors. As such, the spectrum of eigenvalues of $\mathsf{J}(\mathbf{q})$ relates the degree of frustration to the net topology, similar to the spectrum of rigid-unit-mode eigenvalues implicated in structural phase transitions of network materials.\citep{tanRigidUnitMode2023}

Figure~\ref{fig:k_paths} shows the spectrum of eigenvalues of $\mathsf{J}(\mathbf{q})$ for the seven frustrated nets. For strongly-frustrated \textbf{crs} and \textbf{lcv}, the minimum eigenvalue is obtained throughout the Brillouin zone---a three-dimensional degeneracy of propagation vectors---as indicated by flat bands at minimum energy. For weakly-frustrated \textbf{fcu}, a degenerate line of propagation vectors is obtained along $\langle h10 \rangle$ and equivalent directions. We find a similar situation for \textbf{reo}, which has degenerate lines of propagation vectors along $ \langle h00 \rangle$-type directions. This degeneracy is likely responsible for a prevalence of stacking faults that we observed in the \textbf{reo} and \textbf{fcu} ordered states. The degeneracy of \textbf{lcx} and \textbf{lcy} is two-dimensional and hence intermediate between \textbf{crs}/\textbf{lcv} and \textbf{fcu}/\textbf{reo}. For \textbf{lcx}, propagation vectors are degenerate within $\{100\}$-type planes, consistent with linear-chain correlations in real space [Fig.~\ref{fig:heisenberg_summary}]. For \textbf{lcy}, propagation vectors lie on a surface of similar $|\mathbf{q}|$ that includes the observed $(\frac{1}{3}00)$ propagation vector,\citep{isakovFatePartialOrder2008} reminiscent of `spiral spin-liquid' phases induced by competing interactions.\citep{bergmanOrderbydisorderSpiralSpinliquid2007,gaoSpiralSpinliquidEmergence2017} The \textbf{thp} net has two symmetry-unrelated degenerate propagation vectors, $(000)$ and $(\frac{1}{2}\frac{1}{2} \frac{1}{2})$. While this degeneracy is zero-dimensional, there is also significant density of nearly-degenerate wavevectors with slightly higher energies [Fig.~\ref{fig:k_paths}].

\hypertarget{ising}{%
\subsection{Ising}\label{ising}}

Of the seven frustrated nets with Heisenberg interactions, only \textbf{fcu} has a collinear ground state. We
therefore investigated how antiferromagnetic Ising spins will behave on
these nets. We anticipate that the reduction in spin degrees of freedom will destabilise
ordered ground states, as is known for other frustrated nets in which
Heisenberg spins adopt complex ordered states; notably, the Heisenberg
kagome antiferromagnet orders while the Ising equivalent does
not.\citep{kanoAntiferromagnetismKagomeIsing1953, reimersOrderDisorderClassical1993}

\begin{table}
  \renewcommand{\arraystretch}{1.3} 
  \caption{Ground states of the Ising nearest-neighbor antiferromagnets. $f = zJ/3T_\mathrm{N}$ and ${n_\mathrm{flip}}/{n}$ is the flippable spin ratio. \label{tbl:Ising}}
  \begin{tabular}{lrrrl}
    \toprule
    & ${T_\mathrm{N}}/{J}$ & $f$ & ${n_\mathrm{flip}}/{n}$ & Ground state \\
    \midrule
    \textbf{crs}  & <0.01 & >200&  0 & Spin liquid
\citep{liebmannStatisticalMechanicsPeriodic1986} \\
\textbf{fcu} & 1.74\citep{polgreenMonteCarloSimulation1984} & 2.35 &  0 & 2D order$^{\dagger}$
\citep{danielianGroundStateIsing1961} \\
\textbf{lcv} & <0.01 & >130 & 0.44 & Spin
liquid\(^\ast\) \\
\textbf{lcx} & <0.01 & >270 &  0.25 & Spin
liquid\(^\ast\) \\
\textbf{lcy} & <0.01 & >200 & 0.28 & Spin
liquid\(^\ast\)\citep{hopkinsonClassicalAntiferromagnetHyperkagome2007} \\
\textbf{reo} & <0.01 & >270 &  0.24 & Partial order$^{\ddagger}$
\citep{chuiSimpleThreedimensionalIsing1977, reedOrderingSystemFinite1977}\\
\textbf{thp} & <0.01 & >270 &  0.22 & Spin
liquid\(^\ast\) \\
\bottomrule
  \end{tabular}
  \(^\ast\) This work.
  
  \(^\dagger\) In \textbf{fcu}, 2D order occurs at $T=0$ but 3D order occurs at finite temperature as it is favored by fluctuations.\citep{danielianGroundStateIsing1961,polgreenMonteCarloSimulation1984,stubelFinitesizeScalingMonte2018} 
  
  \(^\ddagger\) In \textbf{reo}, partial order with finite entropy occurs at $T=0$\citep{chuiSimpleThreedimensionalIsing1977, reedOrderingSystemFinite1977} but we do not observe an ordering transition at $T/J>0.01$.
  
\end{table}

In moving from the Heisenberg to Ising spin model, we have replaced the
vector spin \(\mathbf{S}\) with a binary variable \(S\). This
Hamiltonian is most likely to be realized in physical systems by binary
degrees of freedom other than magnetic spin, such as site occupancy
or charge,\citep{simonovDesigningDisorderCrystalline2020} since the
designation of a unique magnetic easy axis breaks
the cubic symmetry of the net. We note that
it is possible to designate local easy axes that do maintain the
crystallographic symmetry---e.g., the \textbf{crs} net with ferromagnetic $J$ and local-$\langle 111 \rangle$ easy axes
produces the well-known pyrochlore spin-ice model\citep{bramwellFrustrationIsingtypeSpin1998}---and the effects of such anisotropy in the less well-studied nets could be of interest in future investigations.

\begin{figure}
\centering
\includegraphics{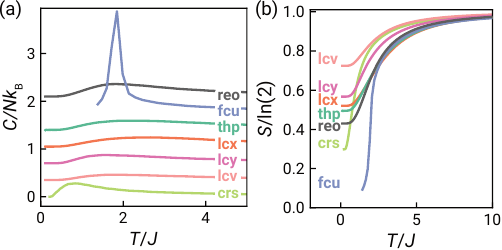}
\caption{(a) Heat capacity (curves offset vertically for clarity) and (b) magnetic entropy for frustrated nets with Ising nearest neighbor interactions derived from Monte Carlo simulations of $10\times10\times10$ supercells ($4\times4\times4$ supercells for \textbf{crs} and \textbf{fcu}).}
\label{fig:ising_heat_cap}
\end{figure}

\begin{figure*}[!ht]
  \centering
  \includegraphics{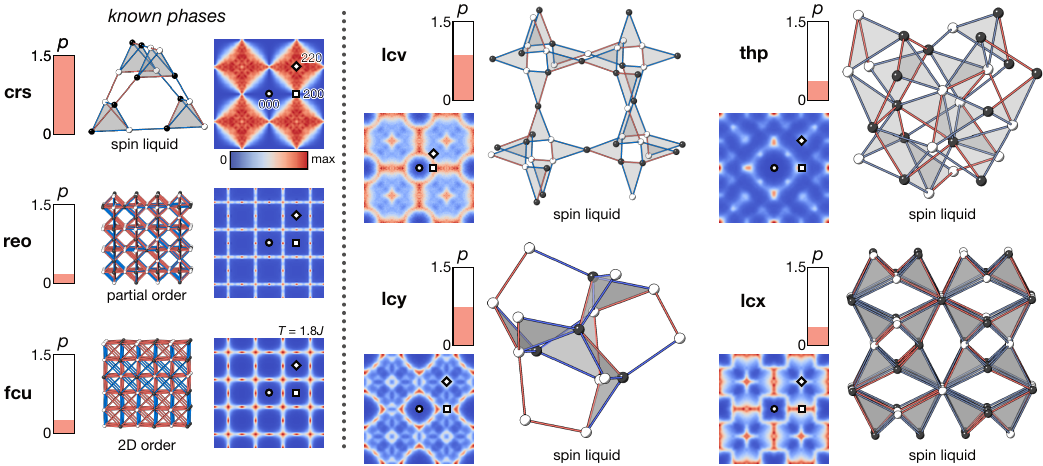}
  \caption{ Summary of the properties of the frustrated nets with Ising spins. The Pauling number $p$ is defined in the text. Structures shown with black and white coloring indicating spin direction, with red bonds indicating ferromagnetic correlation and blue bonds antiferromagnetic correlation. Grey is used to highlight triangles or tetrahedra. Magnetic scattering $S(\mathbf{q})=\left|\sum_{i}\mathbf{S}_{i}\exp(\mathrm{i}\mathbf{q}\cdot\mathbf{r}_{i})\right|^{2}$ is calculated at $T/J = 0.1$ (aside from \textbf{fcu}, which has long-range antiferromagnetic order and hence is shown just above $T_\mathrm{N}$) and is shown in the $(hk0)$ plane, symmetrized in $m\overline{3}m$ symmetry. The $(000)$ position is indicated by a circle, $(200)$ by a square, and $(220)$ by a diamond. A color bar indicating the intensity scale is shown below \textbf{crs}.}
  \label{fig:ising_summary}
\end{figure*}

We find, as anticipated, that reduction of the spin degrees of freedom
reduces the propensity to order in these frustrated systems
[Table~\ref{tbl:Ising}].\citep{liebmannStatisticalMechanicsPeriodic1986}
Out of the seven vertex- and edge-transitive frustrated nets, the
nearest-neighbor global Ising antiferromagnet has been investigated previously for
four. The \textbf{crs} net does not
order.\citep{andersonOrderingAntiferromagnetismFerrites1956,liebmannStatisticalMechanicsPeriodic1986,bramwellFrustrationIsingtypeSpin1998} The \textbf{fcu} net adopts 2D order at $T=0$,\citep{danielianGroundStateIsing1961} but a transition to 3D order occurs at finite temperature due to thermal fluctuations (`order-by-disorder' mechanism).\citep{polgreenMonteCarloSimulation1984,stubelFinitesizeScalingMonte2018} The \textbf{reo} net adopts a partial order with finite
entropy at $T=0$,\citep{chuiSimpleThreedimensionalIsing1977, reedOrderingSystemFinite1977} but its finite-temperature behavior is not known. The \textbf{lcv} net exhibits exponentially-decaying correlations suggesting an absence of order.\citep{hopkinsonClassicalAntiferromagnetHyperkagome2007}
We reproduce these results, finding no long-range order in \textbf{crs}, a clear
peak in heat capacity associated with ordering in \textbf{fcu}, and spin correlations suggestive of 2D order for 
\textbf{reo}, although no phase transition is observed at $T/J>0.01$. Magnetic correlation functions provide further evidence for ordering (or its absence) in each net [Supplementary Figures~2 and 4]. We find that the other four nets \textbf{lcv},
\textbf{lcx}, \textbf{lcy} and \textbf{thp}, which all consist of
corner-sharing triangles alone, do not order down to \(T/J = 0.01\) with Ising spins 
{[}Table \ref{tbl:Ising}, Fig. \ref{fig:ising_heat_cap}{]}. The magnetic
entropy derived from the heat capacity reveals that there is substantial
residual entropy in these four nets {[}Fig. \ref{fig:ising_heat_cap}{]}. A measure indicating the expected level of frustration for these nets with Ising spins can be estimated using the Pauling number,\citep{Overy2015} $p = ( n / 2^{d/2} )$, where $n$ is the number of symmetry equivalent states for each polyhedron and $d$ is the coordination number of each polyhedron. The measure is derived from a generalization and simplification of Pauling's estimate for the configurational entropy of water ice, $S_\mathrm{config} = R \mathrm{ln}(p)$.\citep{Overy2015} The value of $p$ is shown in Figure~\ref{fig:ising_summary}, and correlates with the residual entropy determined through Monte Carlo simulation, with the notable exception of \textbf{crs}, which has significantly less residual magnetic entropy than the Pauling number would predict.

Calculation of the magnetic diffuse scattering pattern for each of these four nets reveals
structured diffuse scattering which resembles the paramagnetic
diffuse scattering observed above \(T_\mathrm{N}\) for the Heisenberg models
{[}Fig.~\ref{fig:ising_summary}{]}. The ratio of accepted to proposed Monte Carlo spin flips for each of
these nets does not tend to zero even at very low temperatures
(\(T/J\sim0.01\)), unlike the known Ising spin liquid on the \textbf{crs}
net. This is due to the presence of flippable spins---spins that have equal
numbers of up and down spin neighbors and hence have no net exchange
field.\citep{moessnerIsingModelsQuantum2001} The acceptance ratio $n_\mathrm{flip}/n$ is
therefore a measure of their concentration, and broadly agrees with the
mean-field predictions (assuming all triangles have two spins `up' and
one `down', or vice versa): \(n_\mathrm{flip}/n = \frac{4}{9} \approx 0.44\) for
two-connected triangles;
\(n_\mathrm{flip}/n = \frac{8}{27} \approx 0.30\) for three-connected triangles,
and \(n_\mathrm{flip}/n = \frac{16}{81} \approx 0.20\) for
four-connected triangles {[}Table \ref{tbl:Ising}{]}. Together with the differences in
structured diffuse scattering and residual entropy, the variation in this
quantity is further evidence that these four
spin liquid states are distinct. The \textbf{lcx}, \textbf{lcy}, and \textbf{thp} nets therefore
host global Ising spin-liquid states that, to the
best of our knowledge, have not previously been investigated.

\hypertarget{conclusion}{%
\section{Conclusion}\label{conclusion}}

In this paper we have enumerated the possible magnetic states of high
symmetry nets with nearest neighbor interactions. We identify that
there are only seven such high-symmetry frustrated nets with nearest-neighbor Heisenberg
antiferromagnetic interactions, and confirm that the only net that hosts a
classical spin liquid ground state is the
well-known \textbf{crs} (pyrochlore) net. We have additionally
identified two nets that have not previously been investigated,
\textbf{lcx} and \textbf{thp}, which have significant frustration as
Heisenberg antiferromagnets and adopt non-collinear ground states.
Further neighbor interactions may destabilise
this 120º order and yield interesting new magnetic physics. We also
predict the form of the magnetic diffuse scattering in the classical spin-liquid regime
(\(J>T>T_\mathrm{N}\)), which suggests significant emergent local order and facilitates experimental
identification of these states using neutron scattering.

We further investigated the behavior of these seven frustrated nets
when decorated with Ising spins. We find that five of these Ising AFMs
show no order down to \(T/J = 0.01\). The four corner-sharing-triangle
nets we identify (\textbf{lcv}, \textbf{lcx}, \textbf{lcy} and
\textbf{thp}) allow significant concentrations of flippable spins
(\(ca. \, 0.3\)), suggestive of significant low temperature dynamics. We
find that the diffuse scattering in these Ising states qualitatively
resembles that of the Heisenberg analogues.

In this study, we have focused on the simplest combinations of
interactions and nets; however, we hope that the discovery of new classical-spin-liquid states even
under these constraints will spur further investigation of these models and potential materials to realize them. We now highlight possible avenues for future work.

First, while we have investigated only the classical models, quantum spins are likely to facilitate the formation of more unusual states; \emph{e.g.}, the classical Heisenberg antiferromagnet
kagome (\textbf{kgm} net) orders magnetically, the \(S=\frac{1}{2}\)
quantum kagome antiferromagnet is considered to have a quantum spin liquid ground
state.\citep{reimersOrderDisorderClassical1993, savaryQuantumSpinLiquids2016}
More complex interactions are likely also to yield interesting states in
the 13 bipartite edge- and vertex-transitive nets; \emph{e.g.}, Kitaev-type bond-directional
interactions are predicted to produce QSL states on the \textbf{srs}
(hyperoctagon or (10,3)a)
net,\citep{obrienClassificationGaplessZ22016} as well on as frustrated
nets.\citep{kimchiKitaevHeisenbergModelsIridates2014} Equally, local
single-ion anisotropy may transform the phases realized on these nets,
even where interactions are ferromagnetic, as is known in the spin-ice Coulomb
phase\citep{bramwellSpinIceState2001} and for the \textbf{lcy} net.\citep{redpathSpinIceTrillium2010} Further neighbor interactions can also `refrustrate' the system and lead to disordered states that differ from the nearest-neighbor model.\citep{baiMagneticExcitationsClassical2019, heneliusRefrustrationCompetingOrders2016}
Recent work has shown that Coulomb phases can
be found on other nets, such as the `octochlore' (\textbf{reo})
phase\citep{szaboFragmentedSpinIce2022} or honeycomb Coulomb
phase.\citep{bentonTopologicalRouteNew2021} The dependence of
these phases on magnetic field, whether an effective transverse
field\citep{moessnerIsingModelsQuantum2001} or external
field,\citep{zhitomirskyHighFieldProperties2005} is also likely to yield
interesting behavior. Finally, the success of MOF chemistry has shown
that topology-guided synthesis of new materials is
feasible,\citep{guillermImportanceHighlyConnected2021} with both
extensive topological
databases\citep{blatovAppliedTopologicalAnalysis2014, OKeeffe2008} and
systematic design rules; \emph{e.g.}, the prediction and potential experimental realization of a Kitaev quantum spin liquid in a metal tetraoxolene honeycomb magnet\citep{Yamada2017, zhangTwoDimensionalCobaltII2024} and the investigation of the centered \textbf{crs} net (or \textbf{crd} net) and discovery of a classical spin liquid phase in \ce{Mn(1,2,3-triazolate)2}.\citep{nutakkiFrustrationCenteredPyrochlore2023} We anticipate that MOFs based on magnetically isotropic ions, such as Mn$^{2+}$, Fe$^{3+}$, or Gd$^{3+}$, may produce the novel Heisenberg phases observed in our simulations, and charge-ordering in mixed-valent MOFs may allow investigation of the new Ising states. We hope therefore that synthetic realizations of these phases will be achievable.

\section*{Acknowledgements}
Work of M.J.C was supported by UKRI (EP/X042782/1). Work of J.A.M.P. was supported by the U.S. Department of Energy, Office of Science, Basic Energy Sciences, Materials Sciences and Engineering Division.

\section*{Author contributions statement}

M.J.C. carried out the simulations. J.A.M.P. wrote the simulation
software. M.J.C. and J.A.M.P. conceived the project and analysed the
simulation results. M.J.C. wrote the paper, with input from J.A.M.P.

\section*{Supporting Information}
Table of all 21 high symmetry nets; figure of all high symmetry nets; radial spin-pair correlation function and nematic correlation functions for each net. Additional data can be freely accessed at doi:10.17639/nott.XXX.

\def\bibsection{\section*{\refname}}
\bibliography{MagNet_CentSci.bib}

\clearpage
\onecolumngrid
\section*{TOC Graphics}
\begin{figure}
  \centering
  \includegraphics{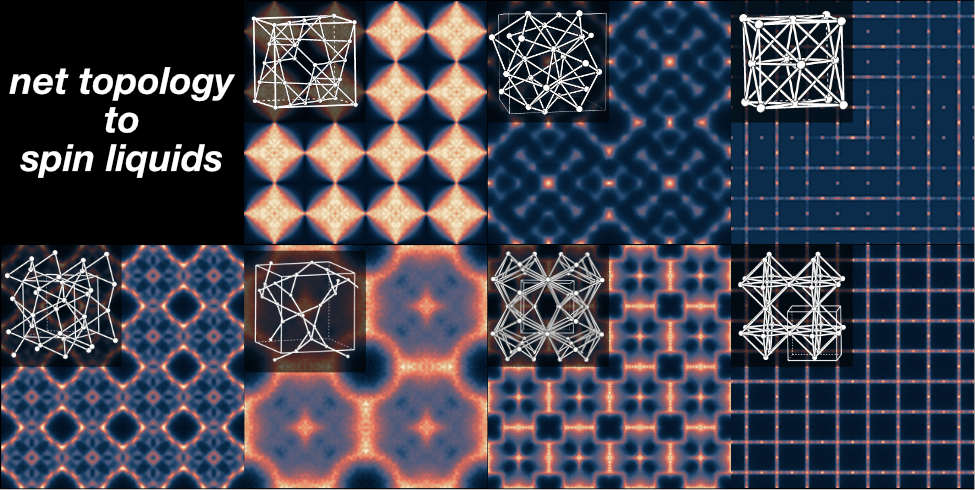}
  \caption{Table of Contents Figure}
  \label{fig:TOC}
  \end{figure}

\section*{Synopsis}
Investigation of high symmetry nets using Monte Carlo simulations uncovered new classical spin liquids with nearest-neighbor antiferromagnetic interactions.

\end{document}